\documentclass[prd,aps,showpacs,nofootinbib,preprint,eqsecnum]{revtex4}
\usepackage[dvips]{graphicx}
\usepackage{color}
\usepackage{amssymb,amsmath}
\usepackage{amsxtra}
\usepackage{epsf}
\usepackage{amssymb}
\usepackage{enumerate}
\usepackage{hhline}
\usepackage{array}
\usepackage{tabularx}
\newcommand{\Eqn}[1]{&\hspace{-0.2em}#1\hspace{-0.2em}&}
\def\e{{\rm e}}
\begin{document}

\title{Cosmological evolution of equation of state for dark energy 
in G-essence models}

\author{
Kazuharu Bamba$^{1, 2,}$\footnote{
E-mail address: bamba@kmi.nagoya-u.ac.jp}, 
Olga Razina$^{2}$, 
Koblandy Yerzhanov$^{2}$ 
and 
Ratbay Myrzakulov$^{2,}$\footnote{
E-mail addresses: rmyrzakulov@gmail.com}}
\affiliation{
$^1$Kobayashi-Maskawa Institute for the Origin of Particles and the
Universe,
Nagoya University, Nagoya 464-8602, Japan\\
$^2$\textit{Eurasian International Center for Theoretical Physics, Dep. Gen. $\&$ Theor.} \\ \textit{Phys., Eurasian National University, Astana 010008, Kazakhstan}}

%
%\date{}
%

\begin{abstract} 
We explore the cosmological evolution of equation of state (EoS) for dark energy in g-essence models, the action of which is described by a function of both the canonical kinetic term of both the scalar and fermionic fields. 
We examine g-essence models with realizing the crossing of the phantom divide line of $w_{\mathrm{DE}}=-1$ 
as well as the models in which the universe always stays in the non-phantom (quintessence) phase ($w_{\mathrm{DE}}>-1$). 
In particular, we find an explicit g-essence model with 
the crossing from the non-phantom phase to the phantom one 
($w_{\mathrm{DE}}<-1$). This transition behavior is consistent with the recent observational data analyses. 
\end{abstract}

\pacs{
%04.50.Kd, 95.36.+x, 98.80.-k
95.36.+x, 98.80.-k
}

\maketitle

%\vspace{2cm} 

%\sloppy

%\tableofcontents

%%%%%%%%%%%%%%%%%%%%%%%%%%%
%%%  Sec. I
%%%%%%%%%%%%%%%%%%%%%%%%%%%
\section{Introduction} 

According to recent cosmological observations such as 
Type Ia Supernovae~\cite{Perlmutter:1998np, Riess:1998cb}, 
cosmic microwave background (CMB) radiation~\cite{Spergel:2003cb, 
Spergel:2006hy, 
Komatsu:2008hk, Komatsu:2010fb, Hinshaw:2012fq}, 
large scale structure (LSS)~\cite{Tegmark:2003ud, Seljak:2004xh}, 
baryon acoustic oscillations (BAO)~\cite{Eisenstein:2005su}, 
and weak lensing~\cite{Jain:2003tba}, 
the expansion of the current universe is accelerating. 

There are two categories to explain the current cosmic acceleration. 
One is to suppose the existence of some unknown matters called dark energy. 
The other is to modify gravity. 
There are a number of candidates for dark energy, e.g., the cosmological constant, scalar fields like quintessence~\cite{Peebles:1987ek, Ratra:1987rm, Chiba:1997ej, Caldwell:1997ii, Zlatev:1998tr} and 
k-essence~\cite{Chiba:1999ka, ArmendarizPicon:1999rj, Garriga:1999vw, ArmendarizPicon:2000dh, ArmendarizPicon:2000ah, dePutter:2007ny}, 
and fluid as a Chaplygin gas~\cite{Kamenshchik:2001cp, Bento:2002ps} 
(for reviews, see, e.g.,~\cite{
Copeland:2006wr, Durrer:2007re, Durrer:2008in, Cai:2009zp, Tsujikawa:2010sc, 
Book-Amendola-Tsujikawa, Li:2011sd, Bamba:2012cp}, 
and for related works, see, for example,~\cite{Yesmakhanova:SMOCAKU, LopezRevelles:2012cg, Bamba:2012gq, Bamba:2012wb, Myrzakulov:KUIBTIC}). 
There also exist several ways of modification of gravity, e.g., 
$F(R)$ gravity (for recent reviews, see, e.g.,~\cite{Nojiri:2010wj, Nojiri:2006ri, Sotiriou:2008rp, Book-Capozziello-Faraoni, Capozziello:2011et, 
DeFelice:2010aj, Clifton:2011jh, Capozziello:2012hm}). 

One of the most important quantities in the studies in terms of the 
property of dark energy is the equation of state (EoS). 
It seem to be suggested by recent various cosmological and astronomical observational data~\cite{Alam:2003fg, Alam:2004jy, Alam:2006kj, Nesseris:2006er, Wu:2006bb, Jassal:2006gf} that 
the crossing of the phantom divide line of $w_{\mathrm{DE}}=-1$ occurred 
in the near past, where $w_{\mathrm{DE}}$ is the EoS for dark energy. 
There are a lot of theoretical attempts and proposals to account for such a 
phantom crossing phenomenon in the framework of both dark energy in general relativity and modified gravity scenarios in the literature (for references, see, e.g., the review articles shown above~\cite{Copeland:2006wr, Durrer:2007re, Durrer:2008in, Cai:2009zp, Tsujikawa:2010sc, Book-Amendola-Tsujikawa, Li:2011sd, Nojiri:2010wj, Nojiri:2006ri, 
Sotiriou:2008rp, Book-Capozziello-Faraoni, Capozziello:2011et, 
DeFelice:2010aj, Clifton:2011jh}.). 

In addition, on the theoretical studies, as phenomenological generalizations of k-essence models, the action of which is written by a function of the canonical kinetic term of a scalar field, 
f-essence and g-essence models have recently been proposed in Ref.~\cite{MR1}. 
The action of f-essence is described by a function of the canonical kinetic term of a fermionic field. On the other hand, the action of g-essence is represented by a function of the canonical kinetic term of a scalar filed as well as that of a fermionic field, which corresponds to a generalized theory constructed by combining k-essence with f-essence. 
In this paper, we concentrate on the evolutions of cosmological quantities, 
especially the EoS for dark energy in g-essence models. 
We analyze the evolutions of cosmological quantities in g-essence models with realizing the crossing of the phantom divide as well as without it. 
In particular, we examine g-essence models in which the universe always 
evolves within the non-phantom (quintessence) phase ($w_{\mathrm{DE}}>-1$). 
In addition, we find an explicit g-essence model with realizing 
the crossing of the phantom divide line of $w_{\mathrm{DE}}=-1$ from the non-phantom phase to the phantom one ($w_{\mathrm{DE}}<-1$). 
It is interesting to emphasize that 
this transition behavior is compatible with the data analyses of the recent cosmological observations~\cite{Alam:2003fg, Alam:2004jy, Alam:2006kj, Nesseris:2006er, Wu:2006bb, Jassal:2006gf}. It has also been investigated that in viable $f(R)$ gravity models, the phantom crossing with the opposite manner (i.e., from the phantom phase to the non-phantom phantom one) can occur (see, e.g.,~\cite{Bamba:2010ws} and the references therein), whereas in $f(T)$ theories, where $T$ is the torsion scalar, the phantom crossing with the same transition direction (i.e., from the non-phantom phase to the phantom one) can happen~\cite{Linder:2010py, Bamba:2010wb, Bamba:2010iw} (for related aspects in $f(T)$ theories, see, e.g.,~\cite{Bamba:2012vg, Jamil:2012nm, Myrzakulov:2012ug}). 

%%%%%%%%
The most significant physical motivation of this work is to illustrate the cosmological evolution of the EoS for dark energy $w_{\mathrm{DE}}$ in g-essence models. This is inspired by the data analysis of recent cosmological and astronomical observations~\cite{Alam:2003fg, Alam:2004jy, Alam:2006kj, Nesseris:2006er, Wu:2006bb, Jassal:2006gf}, which implies that the crossing of the phantom divide happened in the near past. 
On the other hand, 
it is known that in the so-called ``reconstruction''~\cite{Starobinsky:1998fr, Huterer:1998qv, Nakamura:1998mt, Sahni:2006pa, Nojiri:2006gh, Nojiri:2006be, Nojiri:2008fk, Bamba:2008ut, Nojiri:2009kx, Nojiri:2011kd}, we starts with a well motivated theoretical model such as simplicity and/or analogy with some fundamental theory, and by using its form derives an expansion law of the universe to be compared with observational data, or inversely, 
starts with the observational data themselves and try to fit them with a theoretical model. 
Our demonstrations in this work are complementary to such reconstruction 
processes. 
%%%%%%%%

%%%%%
The paper is organized as follows. 
In Sec.\ II, we briefly explain g-essence, k-essence and f-essence 
models. 
In Sec.\ III, we investigate a particular g-essence model and the evolutions of 
cosmological quantities. 
In Sec.\ IV, we explore g-essence models both 
with and without the crossing of the phantom divide. 
%%% 
Finally, conclusions are presented in Sec.\ V. 
%%%%%

%%%%%%%%%%%%%%%%%%%%%%%%%%%
%%%  Sec. II
%%%%%%%%%%%%%%%%%%%%%%%%%%%
\section{Essence models}

In this section, we explain g-essence, k-essence and f-essence 
models. 

%%%%%%%%%%%%%%%%%%%%%%%%%%%
%%%  Sec. II A
%%%%%%%%%%%%%%%%%%%%%%%%%%%
\subsection{g-essence}

The action of g-essence is given by~\cite{MR1}
%reads as  
\begin{equation}
S=\int d^{4}x\sqrt{-g}[R+2K(X, Y, \phi, \psi, \bar{\psi})],
\end{equation} 
where $K$ is some function of its arguments, $\phi$ is a scalar function, $\psi=(\psi_1, \psi_2, \psi_3, \psi_4)^{T}$ is a fermionic function  and $\bar{\psi}=\psi^+\gamma^0$ is its adjoint function.@
Here, the canonical kinetic term for the scalar field $X$ and fermionic field 
$Y$ are given by 
\begin {equation}
X=
%\frac{1}{2}
0.5
g^{\mu\nu}\nabla_{\mu}\phi\nabla_{\nu}\phi\,, \quad 
Y=
%\frac{1}{2}
0.5
i[\bar{\psi}\Gamma^{\mu}D_{\mu}\psi-(D_{\mu}\bar{\psi})\Gamma^{\mu}\psi]\,, 
\end{equation}
%are  the canonical kinetic terms for the scalar and fermionic fields, 
%respectively. 
where 
$\nabla_{\mu}$ and $D_{\mu}$ are the covariant derivatives. 
We note that here the fermionic fields are treated as classically commuting 
fields. 
%Consider  the  homogeneous, isotropic and flat FRW universe filled with g-essen%ce. The metric is given by 

We consider the flat Friedmann-Lema\^{i}tre-Robertson-Walker (FLRW) 
space-time with the metric, 
\begin{equation}
ds^2=dt^2-a^2(dx^2+dy^2+dz^2)\,, 
\end{equation}
where $a(t)$ is the scale factor, 
and the vierbein is chosen to be (see, e.g.,~\cite{Armendariz-Picon})
\begin{equation}
(e_a^\mu)=\mathrm{diag}(1,1/a,1/a,1/a)\,, \quad 
(e^a_\mu)=\mathrm{diag}(1,a,a,a)\,. 
\end{equation}

In the case of the flat FLRW metric (2.3), from the action (2.1) 
the basic equations read~\cite{MR1} 
%corresponding to the action (2.1) look  like 
\begin{eqnarray}
	3H^2-\rho&=&0\,,\\ 
		2\dot{H}+3H^2+p&=&0\,,\\
		K_{X}\ddot{\phi}+(\dot{K}_{X}+3HK_{X})\dot{\phi}-K_{\phi}&=&0\,,\\
		K_{Y}\dot{\psi}+0.5(3HK_{Y}+\dot{K}_{Y})\psi-i\gamma^0K_{\bar{\psi}}&=&0\,,\\ 
K_{Y}\dot{\bar{\psi}}+0.5(3HK_{Y}+\dot{K}_{Y})\bar{\psi}+iK_{\psi}\gamma^{0}&=&0\,,\\
	\dot{\rho}+3H(\rho+p)&=&0\,,
	\end{eqnarray} 
where the kinetic terms, the energy density and pressure 
take the forms 
\begin{equation}
X=0.5\dot{\phi}^2\,, 
\quad  
Y=0.5i(\bar{\psi}\gamma^{0}\dot{\psi}-\dot{\bar{\psi}}\gamma^{0}\psi)\,, 
\quad \rho=2K_{X}X+K_{Y}Y-K\,, \quad
p=K\,.
\end{equation} 
Here, a dot denotes a time derivative of $\partial/\partial t$ and 
$H=\dot{a}/a$ is the Hubble parameter. 
Moreover, the subscription of $K$, e.g., $K_{X} \equiv \partial K/ \partial X$, denotes the derivative of $K$ with respect to $X$. 
Several properties of g-essence have been studied in Refs.~\cite{Razina:2011wv, Razina:2010bj, Kulnazarov:2010an, Yerzhanov:2010mt}. 
%Razina:2011wv, MR4, MR3, MR2}. 
We remark that 
the model (2.1) can also describe 
%admits important two reductions: 
\textit{k-essence} and \textit{f-essence}, as is shown in the following next subsections.

%%%%%%%%%%%%%%%%%%%%%%%%%%%
%%%  Sec. II B
%%%%%%%%%%%%%%%%%%%%%%%%%%%
\subsection{k-essence}
%Let 
We examine the case that the Lagrangian $K$ has the form
\begin{equation}
K=K_1=K_1(X,  \phi)\,.
\end{equation}
Then, the action (2.1) takes the form~\cite{MR1} 
%\cite{}-\cite{}  
\begin {equation}
S=\int d^{4}x\sqrt{-g}[R+2K_1(X,  \phi)]\,,
\end{equation} 
% It 
which corresponds to k-essence~\cite{Chiba:1999ka, ArmendarizPicon:1999rj, Garriga:1999vw, ArmendarizPicon:2000dh, ArmendarizPicon:2000ah, dePutter:2007ny} (for a solvable case, see~\cite{Sharif:2012cu}). 
For the flat FLRW metric (2.3), the basic equations of k-essence 
%look like 
becomes~\cite{Chiba:1999ka, ArmendarizPicon:1999rj, Garriga:1999vw, ArmendarizPicon:2000dh, ArmendarizPicon:2000ah, dePutter:2007ny}  
\begin{eqnarray}
3H^2-\rho&=&0\,,\\ 
2\dot{H}+3H^2+p&=&0\,,\\
K_{X}\ddot{\phi}+(\dot{K}_{X}+3HK_{X})\dot{\phi}-K_{\phi}&=&0\,,\\
\dot{\rho}+3H(\rho+p)&=&0\,.
\end{eqnarray}

%%%%%%%%%%%%%%%%%%%%%%%%%%%
%%%  Sec. II C
%%%%%%%%%%%%%%%%%%%%%%%%%%%
\subsection{f-essence}

Now, we consider the case that the Lagrangian $K$ takes the form
\begin {equation}
K=K_2=K_2(Y,  \psi, \bar{\psi})\,.
\end{equation}
In this case, the action (2.1) is written 
%reads 
as    
\begin {equation}
S=\int d^{4}x\sqrt{-g}[R+2K_2(Y,  \psi, \bar{\psi})]\,.
\end{equation} 
% It 
This is the so-called f-essence~\cite{MR1}. 
In the case of the flat FLRW metric (2.3), the basic equations in f-essence 
have the forms~\cite{MR1}
\begin{eqnarray}
3H^2-\rho&=&0\,,\\ 
2\dot{H}+3H^2+p&=&0\,,\\
K_{Y}\dot{\psi}+0.5(3HK_{Y}+\dot{K}_{Y})\psi-i\gamma^0K_{\bar{\psi}}&=&0\,,\\ 
K_{Y}\dot{\bar{\psi}}+0.5(3HK_{Y}+\dot{K}_{Y})\bar{\psi}+iK_{\psi}\gamma^{0}&=&0\,,\\
\dot{\rho}+3H(\rho+p)&=&0\,,
\end{eqnarray} 
where the energy density and pressure are expressed as 
\begin{equation} 
\rho=K_{Y}Y-K\,, \quad
p=K\,.
\end{equation}
%

%%%%%%%%
It is important to note that 
in addition to the description of 
a spinor field of f-essence as a classical c-number quantity, 
it can also be treated as a Grassmann-valued quantity 
(see, e.g.,~\cite{Damour:2009zc}), or as an operator (``q-number''), as in Ref.~\cite{Damour:2011yk}. 
%%%%%%%%

%%%%%%%%%%%%%%%%%%%%%%%%%%%
%%%  Sec. III
%%%%%%%%%%%%%%%%%%%%%%%%%%%
\section{Particular g-essence model} 

In general, in g-essence models~\cite{MR1} 
%can has the 
equations of motion are very complicated and therefore it is difficult to 
obtain 
%the search  its 
those exact analytical solutions. 
%is a hard job. 
In contrast to such more general and complex cases, 
in order to execute the analyses we here explore 
%modest and work  with 
the following more simple particular model
\begin {equation}
K=\epsilon X+ \sigma Y- V_{1}(\phi)- V_2(u)\,,
\end{equation} 
where $u=\bar{\psi}\psi$, and $\epsilon$ and $\sigma$ are constants. 
%Then 
In this model, 
the equation system (2.9)--(2.14) is described as 
%becomes 
%takes the form
\begin{eqnarray} \label{sys1}
3H^2-\rho&=&0\,,\\ 
2\dot{H}+3H^2+p&=&0\,,\\
\epsilon\ddot{\phi}+3\epsilon H\dot{\phi}+ V_{1\phi}&=&0\,,\\
\sigma\dot{\psi}+\frac{3}{2}\sigma H\psi+i V^{'}_2 \gamma^0 \psi&=&0\,,\\
\sigma\dot{\bar{\psi}}+\frac{3}{2}\sigma H\bar{\psi}-i V^{'}_2\bar{\psi} \gamma^0&=&0\,,\\
\dot{\rho}+3H(\rho+p)&=&0\,, 
\label{sys2}
\end{eqnarray} 
where
\begin{eqnarray}
\rho&=&0.5\epsilon\dot{\phi}^2+V_1+V_2\,, \\ 
p&=&0.5\epsilon\dot{\phi}^2-V_1-V_2+V^{'}_2 u\,.
\end{eqnarray}
Here, a prime denotes the derivative of a quantity with respect of its 
argument such as $V^{'} \equiv \partial V/ \partial u$. 
We note that the contribution of the fermion field is included in 
the potential $V_2 (u)$.

%%%%%%%%%%%%%%%%%%%%%%%%%%%
%%%  Sec. IV
%%%%%%%%%%%%%%%%%%%%%%%%%%%
\section{Cosmological solutions} 

In this section, we analyze cosmological solutions in several examples 
with $V_1=0$ in the particular g-essence model shown in Sec.~III. 
In particular, we examine the cosmological evolutions of 
the energy density $\rho$ and pressure $p$, the deceleration parameter $q \equiv -\ddot{a}/\left(aH^2\right)$, the jerk parameter $j \equiv \dddot{a}/\left(aH^3\right)$~\cite{Sahni:2002fz}, and the EoS $w \equiv p/\rho$ as functions of 
$t$. We study both four examples with and those without the crossing of the phantom divide. 
We note that the subscription $i = 1, 2, \dots, 8$ of $\rho_i$, $p_i$, $q_i$, $j_i$ and $w_i$ denotes the number of four examples in Secs.~IV A 1--4 ($i=1, 2, 3,4$)and those in IV B 1--4 ($i=5, 6, 7, 8$). 
%We also remark that 
In the $\Lambda$ Cold Dark Matter (CDM) model, $q = -1$, $j = 1$ and $w = -1$, 
and hence these quantities denote the deviation of a model from the $\Lambda$CDM model. 
In what follows, we consider the dark energy dominated stage in which 
dark energy (which is, e.g., g-essence in this paper) is dominant over the 
universe, and therefore it can be interpreted that $w \approx w_{\mathrm{DE}}$, where $w_{\mathrm{DE}}$ is the EoS for dark energy. 

It is known that 
in the FLRW background (2.3), 
the effective EoS for the universe is given by~\cite{Nojiri:2010wj, Nojiri:2006ri} 
$
w_{\mathrm{eff}} \equiv p_{\mathrm{eff}}/\rho_{\mathrm{eff}} = 
-1 - 2\dot{H}/\left(3H^2\right)
$ with  
$
\rho_{\mathrm{eff}} \equiv 3H^2/\kappa^2 
$ and 
$
p_{\mathrm{eff}} \equiv -\left(2\dot{H}+3H^2\right)/\kappa^2
$ being the effective energy density and pressure of the universe. 
Here, 
$\rho_{\mathrm{eff}}$ and $p_{\mathrm{eff}}$ correspond to 
the total energy density and pressure of the universe, respectively. 
%%%
If the energy density of dark energy is dominant over 
that of matter completely, $w_{\mathrm{DE}} \approx w_{\mathrm{eff}}$.
%%%%%
In the non-phantom (quintessence) phase, 
$\dot{H} < 0$ and hence $w_\mathrm{eff} >-1$, 
whereas 
in the phantom phase, 
$\dot{H} > 0$ and therefore $w_\mathrm{eff} <-1$. 
Moreover, for $\dot{H} = 0$, 
$w_\mathrm{eff} =-1$, which corresponds to the 
cosmological constant. 
%%%%%
{}From the above considerations, we examine the case that 
$w \approx w_{\mathrm{DE}} \approx w_{\mathrm{eff}}$. 

In case of $V_1=0$, 
the system \eqref{sys1}--\eqref{sys2} becomes
\begin{eqnarray}
3H^2-\rho&=&0\,,\\ 
2\dot{H}+3H^2+p&=&0\,,\\
\ddot{\phi}+3H\dot{\phi}&=&0\,,\\
\sigma\dot{\psi}+\frac{3}{2}\sigma H\psi+i V^{'}_2 \gamma^0 \psi&=&0\,,\\
\sigma\dot{\bar{\psi}}+\frac{3}{2}\sigma H\bar{\psi}-i V^{'}_2\bar{\psi} \gamma^0 &=&0\,,\\
\dot{\rho}+3H(\rho+p)&=&0\,,
\end{eqnarray} 
where
\begin{eqnarray}
\rho&=&0.5\epsilon\dot{\phi}^2+V_2\,, \\ 
p&=&0.5\epsilon\dot{\phi}^2-V_2+V^{'}_2 u\,.
\end{eqnarray}
%Then
As a result, we find 
%have
\begin{equation}
u=\dfrac{c}{a^3},\quad	\dot{\phi}=\dfrac{\phi_0}{a^3}\,, \quad 
V_2=-\frac{0.5\epsilon\phi_0^2}{a^6}+3H^2\,, 
\end{equation}
and
\begin{equation}
\psi_l=\frac{c_l}{a^{1.5}}\e^{iD}\,, \quad  
\psi_k=\frac{c_k}{a^{1.5}}\e^{-iD}\,, 
\end{equation}
where $c$ and $\phi_0$ are constants. 
Here, $l=0, 1$, $k=2, 3$, 
$c_j$ $(j = 0, \dots, 3)$ obey the following condition $c=|c_0|^2+|c_1|^2-|c_2|^2-|c_3|^2$, and
\begin{equation}
D=\frac{1}{c\ \sigma}\left(2\int a^3dH+\phi^2_0\int a^{-3}dt\right)\,.
\end{equation}

%%%%%%%%%%%%%%%%%%%%%%%%%%%
%%%  Sec. IV A
%%%%%%%%%%%%%%%%%%%%%%%%%%%
\subsection{Solutions without the crossing of the phantom divide}
%the phantom divide}

In this subsection, we explore derive the solutions in four examples without the crossing of the phantom divide. 

%%%%%%%%%%%%%%%%%%%%%%%%%%%
%%%  Sec. IV A 1
%%%%%%%%%%%%%%%%%%%%%%%%%%%
\subsubsection{Example 1}

We start from the following expression for the scale factor
\begin{equation}
a=\e^{\dfrac{2}{3}\int\dfrac{dt}{t-0.5(t^2\sin[\frac{1}{t}]+t\cos[\frac{1}{t}]+ Si[ \frac{1}{t}])}}\,,
\end{equation}
where
\begin{equation}
Si[x]=\int_{0}^{x} \frac{\sin[z]}{z}dz\,.
\end{equation}
%Then 
In this case, we obtain
\begin{eqnarray}
	H&=&\dfrac{2}{3[t-0.5(t^2\sin[\frac{1}{t}]+t\cos[\frac{1}{t}]+ Si [\frac{1}{t}])]}\,,\\
	u&=&c\ \e^{-2\int\dfrac{dt}{t-0.5(t^2\sin[\frac{1}{t}]+t\cos[\frac{1}{t}]+ Si [\frac{1}{t}])}}\,, \\
	\dot{\phi}&=&\phi_0 e^{-2\int\dfrac{dt}{t-0.5(t^2\sin[\frac{1}{t}]+t\cos[\frac{1}{t}]+ Si [\frac{1}{t}])}}\,,\\
\psi_l&=&c_l \e^{iD-\int\dfrac{dt}{t-0.5(t^2\sin[\frac{1}{t}]+t\cos[\frac{1}{t}]+ Si[ \frac{1}{t}])}}\,,\\
\psi_k&=&c_k \e^{-iD-\int\dfrac{dt}{t-0.5(t^2\sin[\frac{1}{t}]+t\cos[\frac{1}{t}]+ Si[ \frac{1}{t}])}}\,,
\end{eqnarray}
where 
\begin{align}
D=\frac{1}{c\ \sigma}\int[\dfrac{4(1-t\sin[\frac{1}{t}])}{3[t-0.5(t^2\sin[\frac{1}{t}]+t\cos[\frac{1}{t}]+ Si [\frac{1}{t}])]^2} \e^{2\int\dfrac{dt}{t-0.5(t^2\sin[\frac{1}{t}]+t\cos[\frac{1}{t}]+ Si [\frac{1}{t}])}} +\notag\\
+\phi^2_0 \e^{-2\int\dfrac{dt}{t-0.5(t^2\sin[\frac{1}{t}] +t\cos[\frac{1}{t}]+ Si [\frac{1}{t}])}}]dt\,.
\end{align}
The corresponding potential has the form
%\begin{equation}
\begin{eqnarray}
&&
V_2=-0.5\epsilon\phi_0^2 \e^{-4\int\dfrac{dt} {t-0.5(t^2\sin[\frac{1}{t}]+t\cos[\frac{1}{t}]+ Si [\frac{1}{t}])}} 
\nonumber \\
&&
\hspace{10mm}
{}+\dfrac{4}{3[t-0.5(t^2\sin[\frac{1}{t}]+t\cos[\frac{1}{t}]+ Si [\frac{1}{t}])]^2}\,.
\end{eqnarray}
%\end{equation}
The energy density and pressure are described by 
%have the form
%\begin{equation} 
\begin{eqnarray}
\label{rho1}
\rho_1&=&\dfrac{4}{3[t-0.5(t^2\sin[\frac{1}{t}]+t\cos[\frac{1}{t}]+ Si [\frac{1}{t}])]^2}\,, \nonumber \\
%\quad 
p_1&=&-\dfrac{4t\sin\frac{1}{t}}{3[t-0.5(t^2\sin[\frac{1}{t}]+t\cos[\frac{1}{t}]+ Si [\frac{1}{t}])]^2}\,.
\end{eqnarray}
%\end{equation}
For this example, the EoS 
%equation of state parameter 
and the deceleration parameter have the form
\begin{equation}
%\omega_1
w_1 \equiv \frac{p_1}{\rho_1} 
=-t\sin[\frac{1}{t}]\,,
\end{equation}
and 
\begin{equation}
q_1 \equiv -\frac{\ddot{a}}{aH^2} = 0.5-1.5t\sin[\frac{1}{t}]\,.
\end{equation}

%%%%%% Fig. 1 %%%%%%%%%
\begin{figure}[t]
\centering
\includegraphics[width=0.3 \textwidth, angle=270]{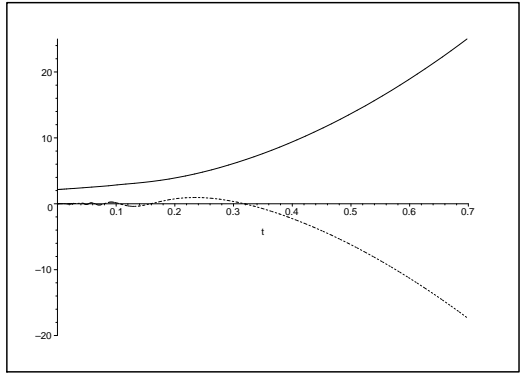}
\caption{$\rho_1$ (solid line) and $p_1$ (dotted line) as a functions of $t$.}
\label{fig:1a}
\end{figure}
%%%%%%%%%%%%%%%%%%%%%%%%

%%%%%% Fig. 2 %%%%%%%%%
\begin{figure}[t]
\centering
\includegraphics[width=0.3 \textwidth, angle=270]{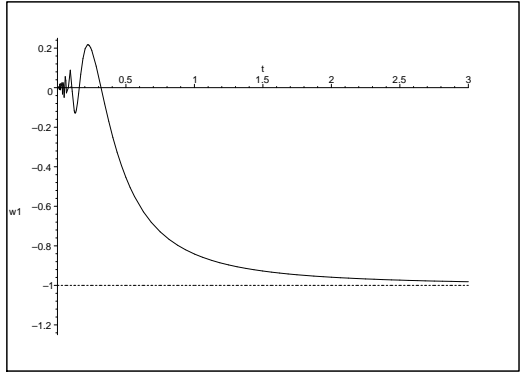}
\caption{$w_1$ as a function of $t$.}
\label{fig:1b}
\end{figure}
%%%%%%%%%%%%%%%%%%%%%%%%

\begin{figure}[t]
\centering
\includegraphics[width=0.30\textwidth, angle=270]{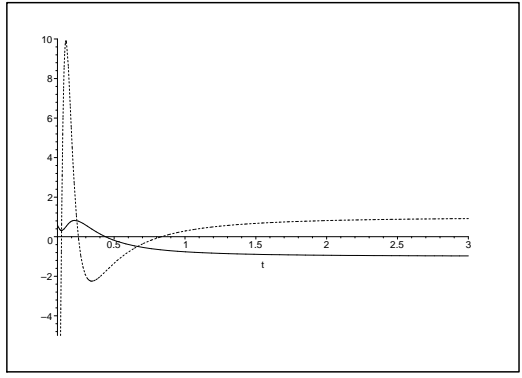}
\caption{$q_1$ (solid line) and $j_1$ (dotted line) as functions of $t$.}
\label{fig:1j}
\end{figure}

In Fig.~\ref{fig:1a}, we show 
%were presented the plots for 
the cosmological evolutions of 
the energy density $\rho_1$ and pressure $p_1$ as the functions of $t$. 
%respectively. 
Furthermore, in Fig.~\ref{fig:1b} we depict the cosmological evolutions of 
the EoS $w_1$ as a function of $t$. {}From Fig.~\ref{fig:1b} we see that 
in this model the crossing of the phantom divide cannot be realized. 
In addition, the jerk parameter is defined as 
%\begin{align}
\begin{eqnarray}
\hspace{-10mm}
j_1 &\equiv&
%\Eqn{\equiv} 
\frac{\dddot{a}}{aH^3} = \frac{\dddot{a}a^2}{\dot{a}^3} 
\nonumber \\
\hspace{-10mm}
&=&
%\Eqn{=} 
1-\frac{9}{8t}\left[(2-3t^3)\sin^2[\frac{1}{t}]+Si[\frac{1}{t}] (t\sin[\frac{1}{t}]-\cos[\frac{1}{t}])+t\cos[\frac{1}{t}](2-\cos[\frac{1}{t}])\right].
%\end{align}
\end{eqnarray}
In Fig.~\ref{fig:1j}, we plot the deceleration parameter $q_1$ and jerk parameter $j_1$ as functions of $t$.

%%%%%% Fig. 3 %%%%%%%%%
%%%%%%%%%%%%%%%%%%%%%%%%

%%%%%%%%%%%%%%%%%%%%%%%%%%%
%%%  Sec. IV A 2
%%%%%%%%%%%%%%%%%%%%%%%%%%%
\subsubsection{Example 2}

We suppose that 
%start from the following expression for 
the scale factor is described by the following expression 
\begin{equation}
a=\e^{\dfrac{2}{3}\int\dfrac{dt}{t-e^{0.5t-1}(t+1)^{0.5(t^2-1)} t^{-0.5t^2}}}.
\end{equation}
Then, we find
\begin{eqnarray}
	H&=&\dfrac{2}{3[t-\e^{0.5t-1}(t+1)^ {0.5(t^2-1)}t^{-0.5t^2}]},\\
	u&=&c\ \e^{-2\int\dfrac{dt}{t-\e^{0.5t-1}(t+1)^{0.5(t^2-1)}t^{-0.5t^2}}}, \\
	\dot{\phi}&=&\phi_0\e^{-2\int\dfrac{dt}{t-\e^{0.5t-1} (t+1)^{0.5(t^2-1)}t^{-0.5t^2}}},\\
\psi_l&=&c_l \e^{iD-\int\dfrac{dt}{t-\e^{0.5t-1}(t+1)^{0.5(t^2-1)} t^{-0.5t^2}}},\\
\psi_k&=&c_k \e^{-iD-\int\dfrac{dt}{t-\e^{0.5t-1}(t+1)^{0.5(t^2-1)} t^{-0.5t^2}}},
\end{eqnarray}
where 
%\begin{align}
\begin{eqnarray}
&&
\hspace{-15mm}
D=\frac{1}{c\ \sigma}\int[\dfrac{4(-t-1+\e^{0.5t-1}(t+1)^{0.5(t^2+1)} t^{-0.5t^2} \ln\dfrac{t+1}{t})}{3(t\sqrt{t+1}-\e^{0.5t-1}(t+1)^ {0.5t^2}t^{-0.5t^2})}\ 
\nonumber \\ 
&&
\hspace{-15mm}
{}\times \e^{2\int\dfrac{dt}{t-\e^{0.5t-1}(t+1)^{0.5(t^2-1)} t^{-0.5t^2}}}
%+\notag\\
+\phi^2_0\int \e^{-2\int\dfrac{dt}{t-\e^{0.5t-1}(t+1)^{0.5(t^2-1)} t^{-0.5t^2}}}]dt.
%\end{align}
\end{eqnarray}
The form of the corresponding potential is given by
\begin{equation}
V_2=-0.5\epsilon\phi_0^2 \e^{-4\int\dfrac{dt}{t-\e^{0.5t-1} (t+1)^{0.5(t^2-1)}t^{-0.5t^2}}}+\dfrac{4}{3[t-\e^{0.5t-1}(t+1)^ {0.5(t^2-1)}t^{-0.5t^2}]^2}.
\end{equation}
The energy density and pressure are expressed as 
\begin{equation} \label{rho1}
	\rho_2=\dfrac{4}{3(t-\e^{0.5t-1}(t+1)^ {0.5(t^2-1)}t^{-0.5t^2})^2},
\quad 
	p_2=-\dfrac{4(1+\frac{1}{t})^t}{3e(t-\e^{0.5t-1}(t+1)^ {0.5(t^2-1)}t^{-0.5t^2})^2}.
\end{equation}
%For this example, 
In case of this example, 
the EoS 
%equation of state parameter 
and the deceleration parameter become 
%have the form
\begin{equation}
	w_2=-e^{-1}(1+\frac{1}{t})^t\,,
\end{equation}
and 
\begin{equation}
	q_2=0.5-1.5e^{-1}(1+\frac{1}{t})^t\,.
\end{equation}

%%%%%% Fig. 4 %%%%%%%%%
\begin{figure}[t]
	\centering
		\includegraphics[width=0.3 \textwidth, angle=270]{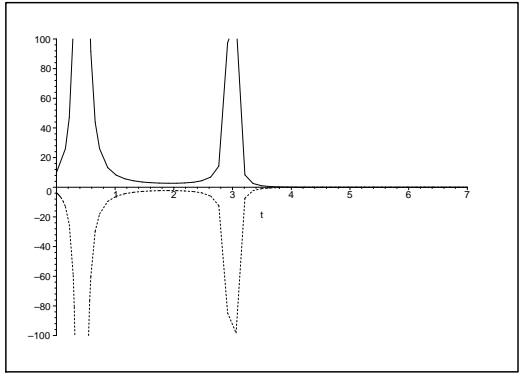}
	\caption{$\rho_2$ (solid line) and $p_2$ (dotted line) 
%Cosmological evolutions of $w_2$ 
as functions of $t$.}
	\label{fig:3a}
\end{figure}
%%%%%%%%%%%%%%%%%%%%%%%%

%%%%%% Fig. 5 %%%%%%%%%
\begin{figure}[t]
	\centering
		\includegraphics[width=0.3 \textwidth, angle=270]{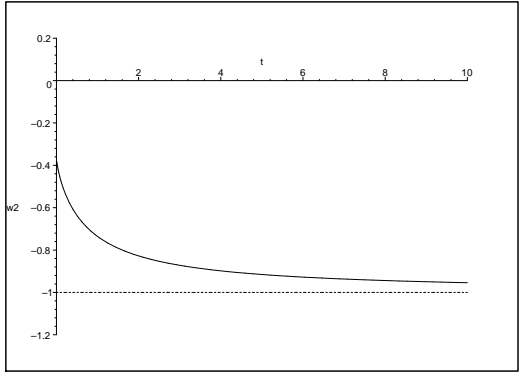}
	\caption{$w_2$ as a function of $t$.}
	\label{fig:3b}
\end{figure}
%%%%%%%%%%%%%%%%%%%%%%%%

\begin{figure}[t]
	\centering
		\includegraphics[width=0.3\textwidth, angle=270]{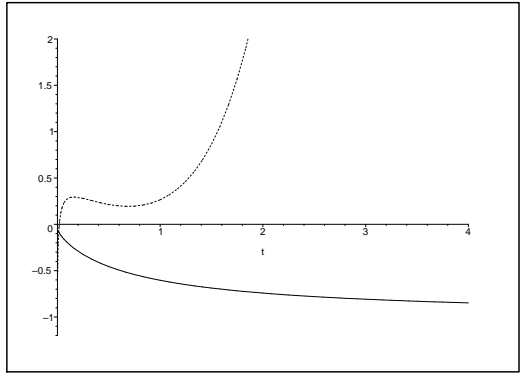}
	\caption{$q_2$ (solid line) and $j_2$ (dotted line) as functions of $t$.}
	\label{fig:2j}
\end{figure}

In Fig.~\ref{fig:3a}, we illustrate the cosmological evolutions of 
%a) were presented the plots for 
the energy density $\rho_2$ and pressure $p_2$ as functions of $t$. 
%, respectively. 
In Fig.~\ref{fig:3b}, we show  
%b) 
the cosmological evolutions of the EoS $w_2$ as a function of $t$. 
It follows from Fig.~\ref{fig:3b} that 
the crossing of the phantom divide cannot occur in this 
example.
%
%Moreover, the jerk parameter is given by 
%\begin{align}
%	j_2=\frac{\dddot{a}}{aH^3}=\frac{\dddot{a}a^2}{\dot{a}^3}\,.
%\end{align}
%
In Fig.~\ref{fig:2j}, we plot 
% were presented the plots for 
the deceleration parameter $q_2$ and jerk parameter $j_2$\footnote{In Fig.~\ref{fig:2j}, we have plotted the behavior of $j_2$ numerically because the analytic expression is too complicated. This is the same as for $j_3$ in Fig.~\ref{fig:3jb} in Sec.~IV A 3.} as functions of $t$. 
%, respectively

%%%%%% Fig. 6 %%%%%%%%%
%%%%%%%%%%%%%%%%%%%%%%%%

%%%%%%%%%%%%%%%%%%%%%%%%%%%
%%%  Sec. IV A 3
%%%%%%%%%%%%%%%%%%%%%%%%%%%
\subsubsection{Example 3}

We provide that 
%start from the following expression for 
the scale factor is expressed as 
\begin{equation}
a=\e^{\dfrac{2}{3}\int\dfrac{dt}{t-A^{0.5A^2}\e^{A(0.5t-1)}(t+A)^ {0.5(t^2-A^2)}t^{-0.5t^2}}}\,,
\end{equation}
where $A$ is a constant. {}From this expression, we have 
%Then we obtain
\begin{eqnarray}
	H&=&\dfrac{2}{3[t-A^{0.5A^2}\e^{A(0.5t-1)}(t+A)^ {0.5(t^2-A^2)}t^{-0.5t^2}]}\,,\\
	u&=&c\ \e^{-2\int\dfrac{dt}{t-A^{0.5A^2}\e^{A(0.5t-1)}(t+A)^ {0.5(t^2-A^2)}t^{-0.5t^2}}}\,, \\
	\dot{\phi}&=&\phi_0  \e^{-2\int\dfrac{dt}{t-A^{0.5A^2}\e^{A(0.5t-1)}(t+A)^ {0.5(t^2-A^2)}t^{-0.5t^2}}}\,,\\
\psi_l&=&c_l\e^{iD-\int\dfrac{dt}{t-A^{0.5A^2}\e^{A(0.5t-1)}(t+A)^ {0.5(t^2-A^2)}t^{-0.5t^2}}}\,,\\
\psi_k&=&c_k \e^{-iD-\int\dfrac{dt}{t-A^{0.5A^2}\e^{A(0.5t-1)}(t+A)^ {0.5(t^2-A^2)}t^{-0.5t^2}}}\,,
\end{eqnarray}
where 
\begin{align}
D=\frac{1}{c\ \sigma}\int[ 
\dfrac{4(-1+A^{0.5A^2}\e^{A(0.5t-1)}(t+A)^ {0.5(t^2-A^2)}t^{-0.5t^2+1}\ln \dfrac{t+A}{t})}{3(t-A^{0.5A^2}\e^{A(0.5t-1)}(t+A)^ {0.5(t^2-A^2)}t^{-0.5t^2})^2}\times\notag\\ \times\ \e^{2\int\dfrac{dt}{t-A^{0.5A^2}\e^{A(0.5t-1)}(t+A)^ {0.5(t^2-A^2)}t^{-0.5t^2}}}+\notag\\
+\phi^2_0\ \e^{-2\int\dfrac{dt}{t-A^{0.5A^2}\e^{A(0.5t-1)}(t+A)^ {0.5(t^2-A^2)}t^{-0.5t^2}}}] dt\,.
\end{align}
The corresponding potential becomes 
%has the form
\begin{align}
	V_2=-0.5\epsilon\phi_0^2 \e^{-4\int\dfrac{dt}{t-A^{0.5A^2}\e^{A(0.5t-1)}(t+A)^ {0.5(t^2-A^2)}t^{-0.5t^2}}}+\notag\\
	+\dfrac{4}{3[t-A^{0.5A^2}\e^{A(0.5t-1)}(t+A)^ {0.5(t^2-A^2)}t^{-0.5t^2}]^2}\,. 
\end{align}
The energy density and pressure are given by 
%have the form
%\begin{equation} 
\begin{eqnarray}
	\rho_3\Eqn{=}\dfrac{4}{3[t-A^{0.5A^2}\e^{A(0.5t-1)}(t+A)^ {0.5(t^2-A^2)}t^{-0.5t^2}]^2}\,,
\label{rho1} \\
%\end{equation}
%\begin{equation}
p_3\Eqn{=}-\dfrac{4(1+\frac{A}{t})^t}{3e^A[t-A^{0.5A^2}\e^{A(0.5t-1)}(t+A)^ {0.5(t^2-A^2)}t^{-0.5t^2}]^2}\,.
%\end{equation}
\end{eqnarray}
%For 
In this example, the EoS 
%equation of state parameter 
and the deceleration parameter are written as 
%have the form
\begin{equation}
		w_3=-\e^{-A}(1+\frac{A}{t})^t
\end{equation}
and 
\begin{equation}
	q_3=0.5-1.5\e^{-A}(1+\frac{A}{t})^t\,, 
\end{equation}
respectively.

%%%%%% Fig. 7 %%%%%%%%%
\begin{figure}[t]
	\centering
		\includegraphics[width=0.3 \textwidth, angle=270]{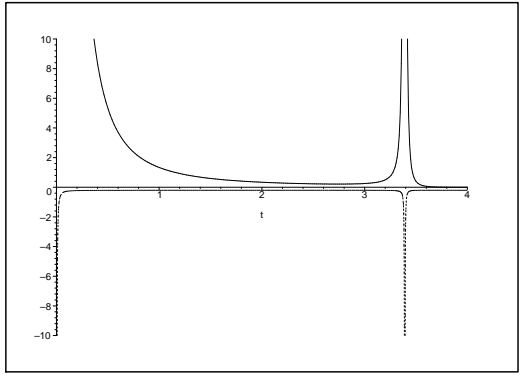}
	\caption{$\rho_3$ (solid line) and $p_3$ (dotted line) as functions of $t$ for $A=10$.}
	\label{fig:4a}
\end{figure}
%%%%%%%%%%%%%%%%%%%%%%%%

%%%%%% Fig. 8 %%%%%%%%%
\begin{figure}[t]
	\centering
		\includegraphics[width=0.3 \textwidth, angle=270]{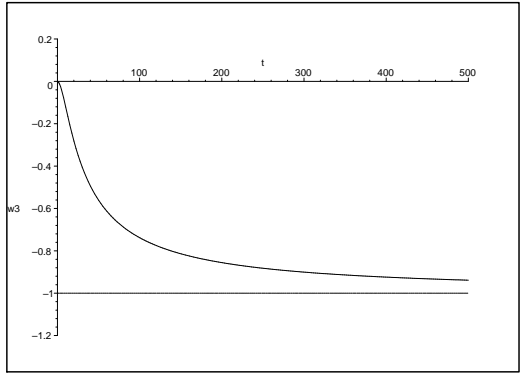}
	\caption{$w_3$ as a function of $t$ 
for $A=10$.
%Legend is the same as in Fig.~\ref{fig:4a}.
}
	\label{fig:4b}
\end{figure}
%%%%%%%%%%%%%%%%%%%%%%%%

\begin{figure}[t]
	\centering
		\includegraphics[width=0.3 \textwidth, angle=270]{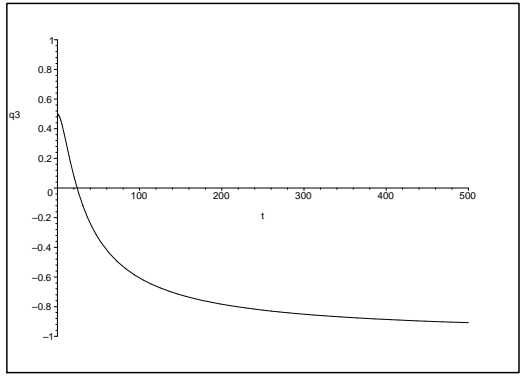}
	\caption{$q_3$ (solid line) as a function of $t$.}
	\label{fig:3ja}
\end{figure}

\begin{figure}[t]
	\centering
		\includegraphics[width=0.3 \textwidth, angle=270]{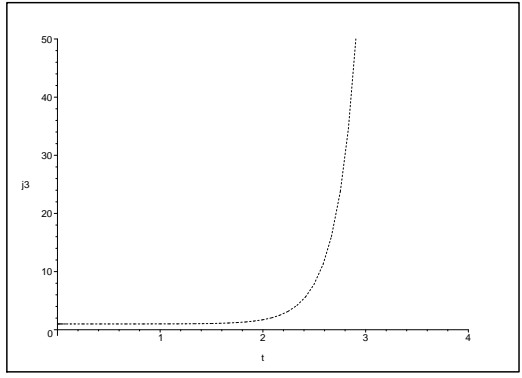}
	\caption{$j_3$ (dotted line) as a function of $t$.}
	\label{fig:3jb}
\end{figure}

In Fig.~\ref{fig:4a}, we plot the cosmological evolutions of the energy density $\rho_3$ and pressure $p_3$ as functions of $t$ for $A=10$.
In Fig.~\ref{fig:4b}, we display the cosmological evolution of the EoS $w_3$ as a function of $t$ for $A=10$. We find from Fig.~\ref{fig:4b} that 
in this model the crossing of the phantom divide cannot happen. 
%
%The jerk parameter
%\begin{align}
%	j_3=\frac{\dddot{a}}{aH^3}=\frac{\dddot{a}a^2}{\dot{a}^3}\,.
%\end{align}
%
In Fig.~\ref{fig:3ja}, we show the cosmological evolution of the deceleration parameter $q_3$ as a function of $t$. 
In Fig.~\ref{fig:3jb}, we illustrate the cosmological evolution of the jerk parameter $j_3$\footnote{See the footnote in Sec.~IV A 2.} as a function of $t$.

%%%%%% Fig. 9 %%%%%%%%%
%%%%%%%%%%%%%%%%%%%%%%%%

%%%%%% Fig. 10 %%%%%%%%%
%%%%%%%%%%%%%%%%%%%%%%%%

%%%%%%%%%%%%%%%%%%%%%%%%%%%
%%%  Sec. IV A 4
%%%%%%%%%%%%%%%%%%%%%%%%%%%
\subsubsection{Example 4}

We consider that the scale factor is written by 
%start from 
%the following expression for 
\begin{equation}
a=(a_0+b\cosh[kt]+d\sinh[kt])^{\frac{1}{3}},
\end{equation}
where $a_0$, $b$, $d$ and $k$ are constants. 
In this case, we acquire 
\begin{eqnarray}
H&=&\frac{k(b\sinh[kt]+d\cosh[kt])}{3(a_0+b\cosh[kt]+d\sinh [kt])}\,,\\
u&=&\frac{c}{a_0+b\cosh[kt]+d\sinh[kt]}\,, \\
\phi&=&\frac{2\phi_0^2} {k\sqrt{b^2-d^2-a_0^2}}\arctan\left(\frac{(b-a_0)\tanh[0.5kt]+d}{\sqrt{b^2-d^2-a_0^2}}\right)\,,\\
\psi_l&=&\frac{c_l}{(a_0+b\cosh[kt]+d\sinh[kt])^{2}}\e^{iD}\,,\\
\psi_k&=&\frac{c_k}{(a_0+b\cosh[kt]+d\sinh[kt])^{2}}\e^{-iD}\,,
\end{eqnarray}
where 
%\begin{align}
\begin{eqnarray}
&&
D=\frac{2k}{3c\ \sigma}\left[\frac{2(b^2-d^2-a_0^2+\dfrac{3\phi_0^2}{2k^2})}{\sqrt{b^2-d^2-a_0^2}}\arctan\left(\frac{(b-a_0)\tanh[0.5kt]+d}{\sqrt{b^2-d^2-a_0^2}}\right) \right. \nonumber \\
&& \left. 
\hspace{20mm}
{}+a_0 \ln\left(\frac{\tanh[0.5kt+1]}{\tanh[0.5kt-1]}\right)\right]\,.
\end{eqnarray}
%\end{align}
The corresponding potential is given by 
%has the form
\begin{equation}
V_2=-\frac{0.5\epsilon\phi_0^2}{(a_0+b\cosh[kt]+d\sinh[kt])^2}+\frac{k^2}{3}\left( \frac{b\sinh[kt]+d\cosh[kt]}{a_0+b\cosh[kt]+d\sinh [kt]}\right)^2\,.
\end{equation}
The energy density and pressure have the forms 
%\begin{equation} 
\begin{eqnarray}
	\rho_4\Eqn{=}\frac{k^2}{3}\left( \frac{b\sinh[kt]+d\cosh[kt]}{a_0+b\cosh[kt]+d\sinh [kt]}\right)^2\,,
\label{rho1} \\
%\end{equation}
%\begin{equation}
p_4\Eqn{=}-\frac{k^2}{3}\frac{(b\sinh [kt]+d\cosh [kt])^2-2[a_0(b\cosh[kt]+d\sinh [kt])+b^2-d^2]}{(a_0+b\cosh [kt]+d\sinh [kt])^2}\,.
\end{eqnarray}
%\end{equation}
For this example, 
the EoS and the deceleration parameter become 
\begin{equation}
	\omega_4=-1-\frac{2[a_0(b\cosh[kt]+d\sinh[kt])+b^2-d^2]}{(b\sinh [kt]+d\cosh [kt])^2}
\end{equation}
and 
\begin{equation}
	q_4=-1-\frac{3[a_0(b\cosh[kt]+d\sinh[kt])+b^2-d^2]}{(b\sinh [kt]+d\cosh [kt])^2}\,,
\end{equation}
respectively.

%%%%%% Fig. 11 %%%%%%%%%
\begin{figure}[t]
	\centering
		\includegraphics[width=0.3 \textwidth, angle=270]{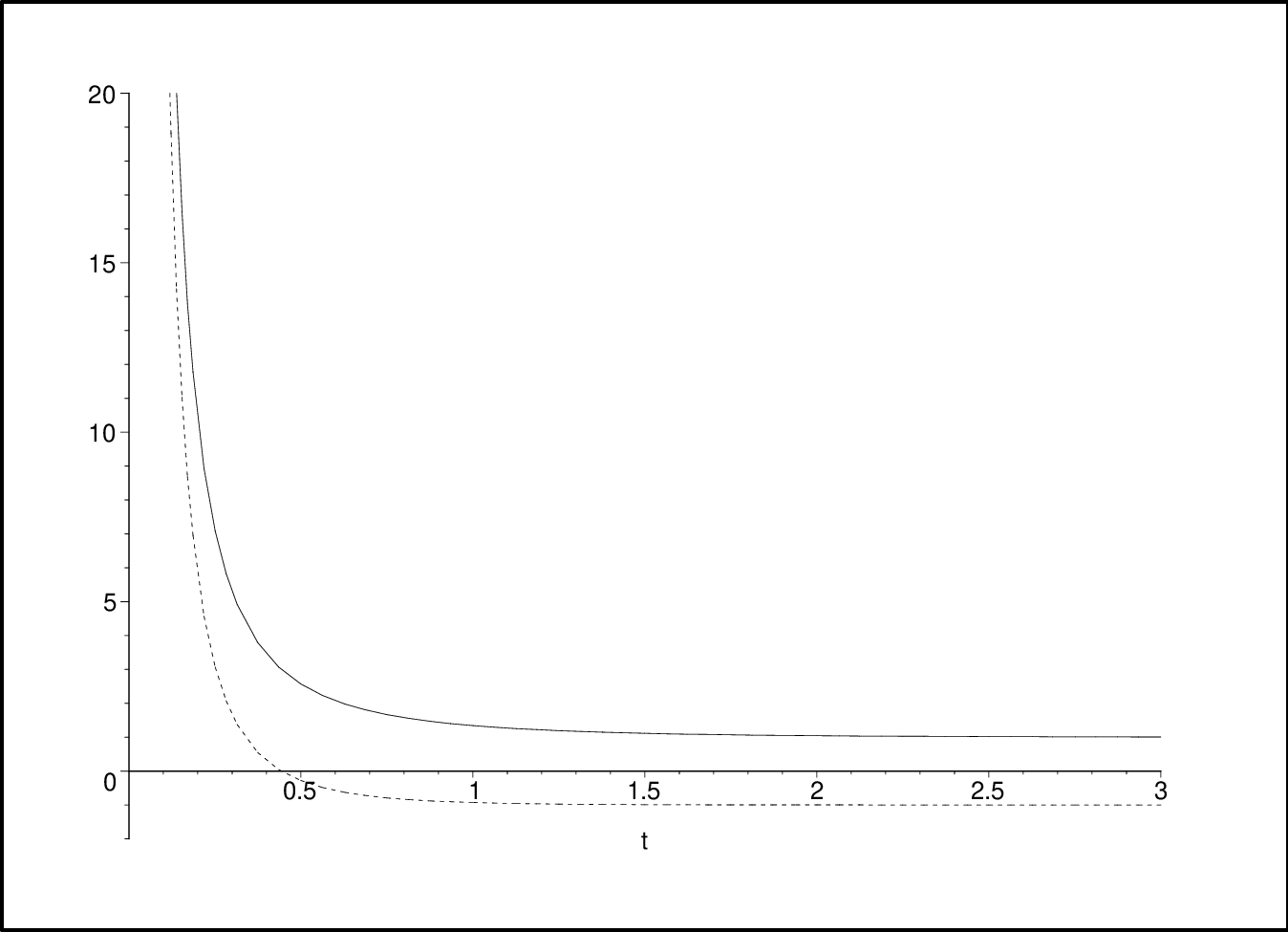}
	\caption{$\rho_4$ (solid line) and $p_4$ (dotted line) as functions of $t$.}
	\label{fig:7a}
\end{figure}
%%%%%%%%%%%%%%%%%%%%%%%%

%%%%%% Fig. 12 %%%%%%%%%
\begin{figure}[t]
	\centering
		\includegraphics[width=0.3 \textwidth, angle=270]{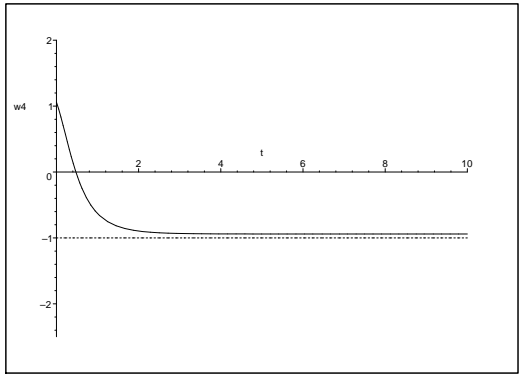}
	\caption{$w_4$ as a function of $t$ for $a_0=-1/2$, $b=1/2$, $d=1$ and $k=\sqrt{3}$.}
	\label{fig:7b}
\end{figure}
%%%%%%%%%%%%%%%%%%%%%%%%

\begin{figure}[t]
	\centering
		\includegraphics[width=0.3 \textwidth, angle=270]{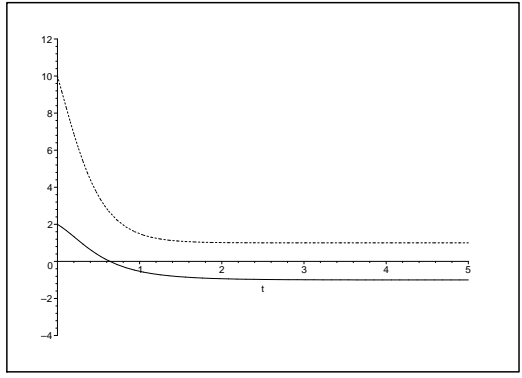}
	\caption{$q_4$ (solid line) and $j_4$ (dotted line) as functions of $t$.}
	\label{fig:4j}
\end{figure}

In Fig.~\ref{fig:7a}, we depict the cosmological evolutions of the energy density $\rho_4$ and pressure $p_4$ as functions of $t$. 
In Fig.~\ref{fig:7b}, we show the cosmological evolution of the EoS $w_4$ as a function of $t$. {}From Fig.~\ref{fig:7b}, 
we understand that also 
in this example the crossing of the phantom divide cannot be realized. 
Moreover, the jerk parameter is described by 
\begin{align}
	j_4=\frac{9a^2_0-10b^2+9d^2+(b^2+d^2)\cosh^2[kt]+bd\sinh[2kt]}{(b\sinh[kt]+d\cosh[kt])^2}\,.
\end{align}
In Fig.~\ref{fig:4j}, we illustrate the cosmological evolutions of the deceleration parameter $q_4$ and jerk parameter $j_4$ as functions of $t$.

%%%%%% Fig. 13 %%%%%%%%%
%%%%%%%%%%%%%%%%%%%%%%%%

In summary, 
from Figs.~\ref{fig:1b}, \ref{fig:3b}, \ref{fig:4b} and \ref{fig:7b} 
we see that 
in the above four models, the universe is always in the non-phantom 
(quintessence) phase (in which $w > -1$) and therefore 
the crossing of the phantom divide cannot be realized.

%%%%%%%%%%%%%%%%%%%%%%%%%%%
%%%  Sec. IV 2
%%%%%%%%%%%%%%%%%%%%%%%%%%%
\subsection{Solutions with the crossing of the phantom divide}
%the phantom divide}

Next, in this subsection 
we investigate the solutions in four examples with realizing the crossing of the phantom divide.

%%%%%%%%%%%%%%%%%%%%%%%%%%%
%%%  Sec. IV B 1
%%%%%%%%%%%%%%%%%%%%%%%%%%%
\subsubsection{Example 1}

We consider that 
%start from the following expression for 
the scale factor is described by 
\begin{equation}
	a=\e^{\dfrac{2}{3}\int\dfrac{dt}{t-\int t \tan [\frac{1}{t}]dt}}\,.
\end{equation}
In this example, we acquire 
\begin{eqnarray}
H&=&\dfrac{2}{3(t-\int t \tan [\frac{1}{t}]dt)}\,,\\
u&=&c\ \e^{-2\int\dfrac{dt}{t-\int t \tan [\frac{1}{t}]dt}}\,, \\
\dot{\phi}&=&\phi_0\e^{-2\int\dfrac{dt}{t-\int t \tan [\frac{1}{t}]dt}}\,,\\
\psi_l&=&c_l \e^{iD-\int\dfrac{dt}{t-\int t \tan [\frac{1}{t}]dt}}\,,\\
\psi_k&=&c_k \e^{-iD-\int\dfrac{dt}{t-\int t \tan [\frac{1}{t}]dt}}\,,
\end{eqnarray}
where 
\begin{align}
D=\frac{1}{c\ \sigma}\int\left[\dfrac{4(t\tan[\frac{1}{t}]-1)}{3(t-\int t \tan [\frac{1}{t}]dt)^2}\ \e^{2\int\dfrac{dt}{t-\int t \tan [\frac{1}{t}]dt}}+\phi^2_0\ \e^{-2\int\dfrac{dt}{t-\int t \tan [\frac{1}{t}]dt}}\right]dt\,.
\end{align}
The form of the corresponding potential reads 
\begin{equation}
V_2=-0.5\epsilon\phi_0^2\e^{-4\int\dfrac{dt}{t-\int t \tan [\frac{1}{t}]dt}}+\dfrac{4}{3(t-\int t \tan [\frac{1}{t}]dt)^2}\,.
\end{equation}
The energy density and pressure are given by 
\begin{equation} \label{rho1}
	\rho_5=\dfrac{4}{3(t-\int t \tan [\frac{1}{t}]dt)^2}\,,
\quad 
p_5=-\dfrac{4t\tan[\frac{1}{t}]}{3(t-\int t \tan[ \frac{1}{t}]dt)^2}\,.
\end{equation}
For this example, the EoS and the deceleration parameter are written by 
\begin{equation}
w_5=-t\tan[\frac{1}{t}]
\end{equation}
and 
\begin{equation}
	q_5=0.5-1.5t\tan[\frac{1}{t}]\,, 
\end{equation}
respectively.

%%%%%% Fig. 14 %%%%%%%%%
\begin{figure}[t]
	\centering
		\includegraphics[width=0.3 \textwidth, angle=270]{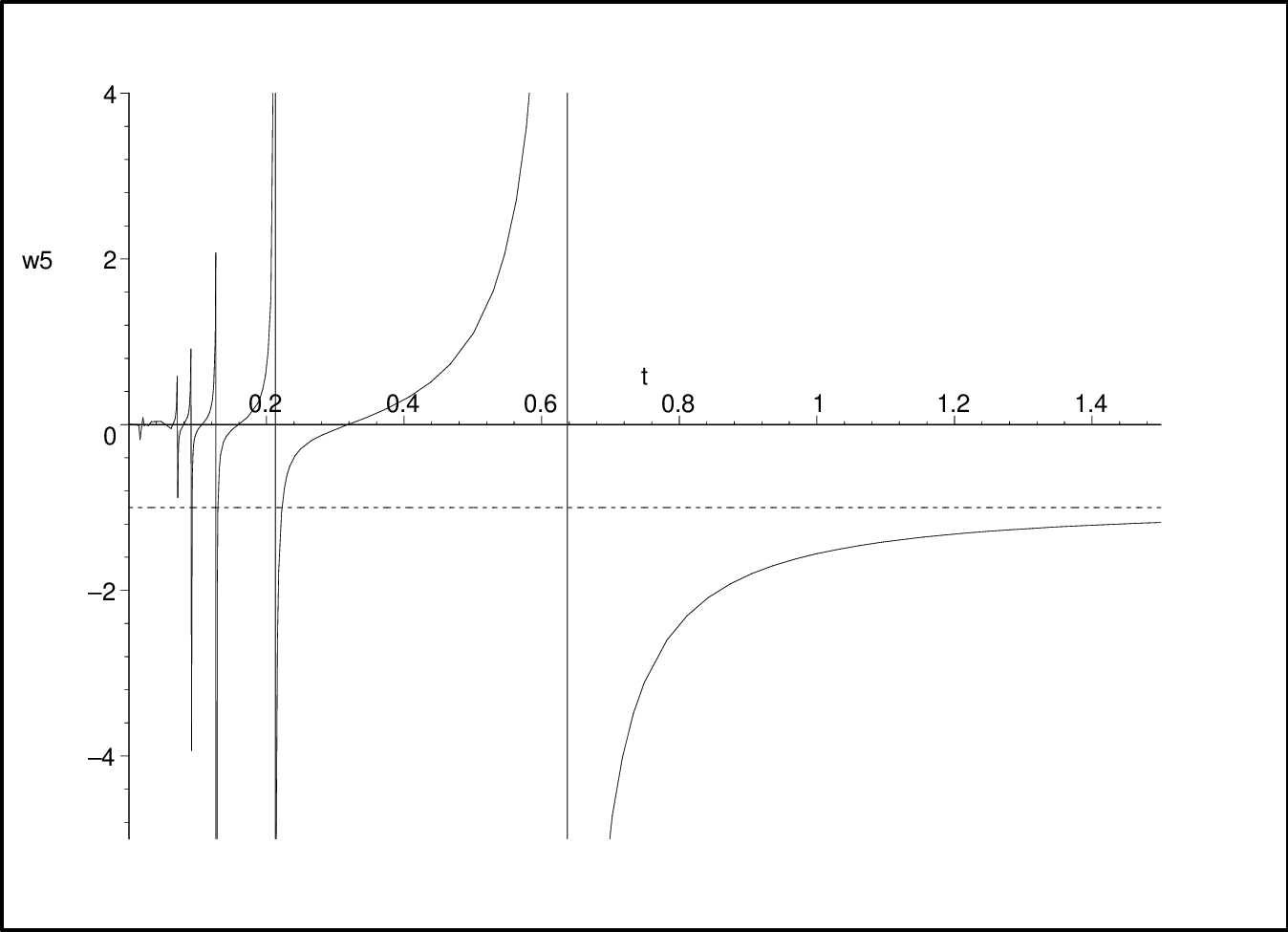}
	\caption{$w_5$ as a function of $t$.}
	\label{fig:2}
\end{figure}
%%%%%%%%%%%%%%%%%%%%%%%%

In Fig.~\ref{fig:2}, we display the EoS $w_5$ as a function of $t$. 
{}From Fig.~\ref{fig:2}, 
we see that in this model the crossing of the phantom divide can be realized. 
In addition, the jerk parameter is described by 
\begin{align}
	j_5=-\frac{1}{4t}\left[5t+9t\tan[\frac{1}{t}]\left(t-(2t^2-1)\tan[\frac{1}{t}]\right)+9\left(t\tan[\frac{1}{t}]-\tan^2[\frac{1}{t}]-1\right)\int t\tan[\frac{1}{t}]dt\right].
\end{align}

%%%%%%%%%%%%%%%%%%%%%%%%%%%
%%%  Sec. IV B 2
%%%%%%%%%%%%%%%%%%%%%%%%%%%
\subsubsection{Example 2}

We take 
%start from the following expression for 
the scale factor as 
\begin{equation}
a=\e^{\dfrac{2}{3}\int\dfrac{dt}{t-0.5(t^2\sin[\frac{1}{t}]+t\cos[\frac{1}{t}])}}\,.
\end{equation}
%Then we obtain
In this case, we find 
\begin{eqnarray}
	H&=&\dfrac{2}{3[t-0.5(t^2\sin[\frac{1}{t}]+t\cos[\frac{1}{t}])]}\,,\\
	u&=&c\ \e^{-2\int\dfrac{dt}{t-0.5(t^2\sin[\frac{1}{t}]+t\cos[\frac{1}{t}])}}\,, \\
	\dot{\phi}&=&\phi_0 \e^{-2\int\dfrac{dt}{t-0.5(t^2\sin[\frac{1}{t}]+t\cos[\frac{1}{t}])}}\,,\\
\psi_l&=&c_l\e^{iD-\int\dfrac{dt}{t-0.5(t^2\sin[\frac{1}{t}]+t\cos[\frac{1}{t}])}}\,,\\
\psi_k&=&c_k\e^{-iD-\int\dfrac{dt}{t-0.5(t^2\sin[\frac{1}{t}]+t\cos[\frac{1}{t}])}}\,,
\end{eqnarray}
where 
\begin{align}
D=\frac{1}{c\ \sigma}\int[ \dfrac{4(1-[t+0.5t^{-1}]\sin[\frac{1}{t}])}{3(t-0.5(t^2\sin[\frac{1}{t}]+t\cos[\frac{1}{t}]))^2}\ \e^{2\int\dfrac{dt}{t-0.5(t^2\sin[\frac{1}{t}]+t\cos[\frac{1}{t}])}} +\notag\\
+\phi^2_0\int \e^{-2\int\dfrac{dt}{t-0.5(t^2\sin[\frac{1}{t}] +t\cos[\frac{1}{t}])}}]dt\,.
\end{align}
The corresponding potential is given by 
\begin{equation}
	V_2=-0.5\epsilon\phi_0^2 \e^{-4\int\dfrac{dt} {t-0.5(t^2\sin[\frac{1}{t}]+t\cos[\frac{1}{t}])}}    +\dfrac{4}{3[t-0.5(t^2\sin[\frac{1}{t}]+t\cos[\frac{1}{t}])]^2}\,.
\end{equation}
The forms of the energy density and pressure become 
\begin{equation} \label{rho1}
\rho_6=\dfrac{4}{3[t-0.5(t^2\sin[\frac{1}{t}]+t\cos[\frac{1}{t}])]^2}\,,
\quad 
p_6=-\dfrac{4(t+0.5t^{-1})\sin[\frac{1}{t}]}{3[t-0.5
(t^2\sin[\frac{1}{t}]+t\cos[\frac{1}{t}])]^2}\,.
\end{equation} 
For this example, the EoS and the deceleration parameter are given by 
\begin{equation}
	w_6=-(t+0.5t^{-1})\sin[\frac{1}{t}]
\end{equation}
and 
\begin{equation}
	q_6=0.5-1.5(t+0.5t^{-1})\sin[\frac{1}{t}]\,,
\end{equation}
respectively.

%%%%%% Fig. 15 %%%%%%%%%
\begin{figure}[t]
	\centering
		\includegraphics[width=0.3 \textwidth, angle=270]{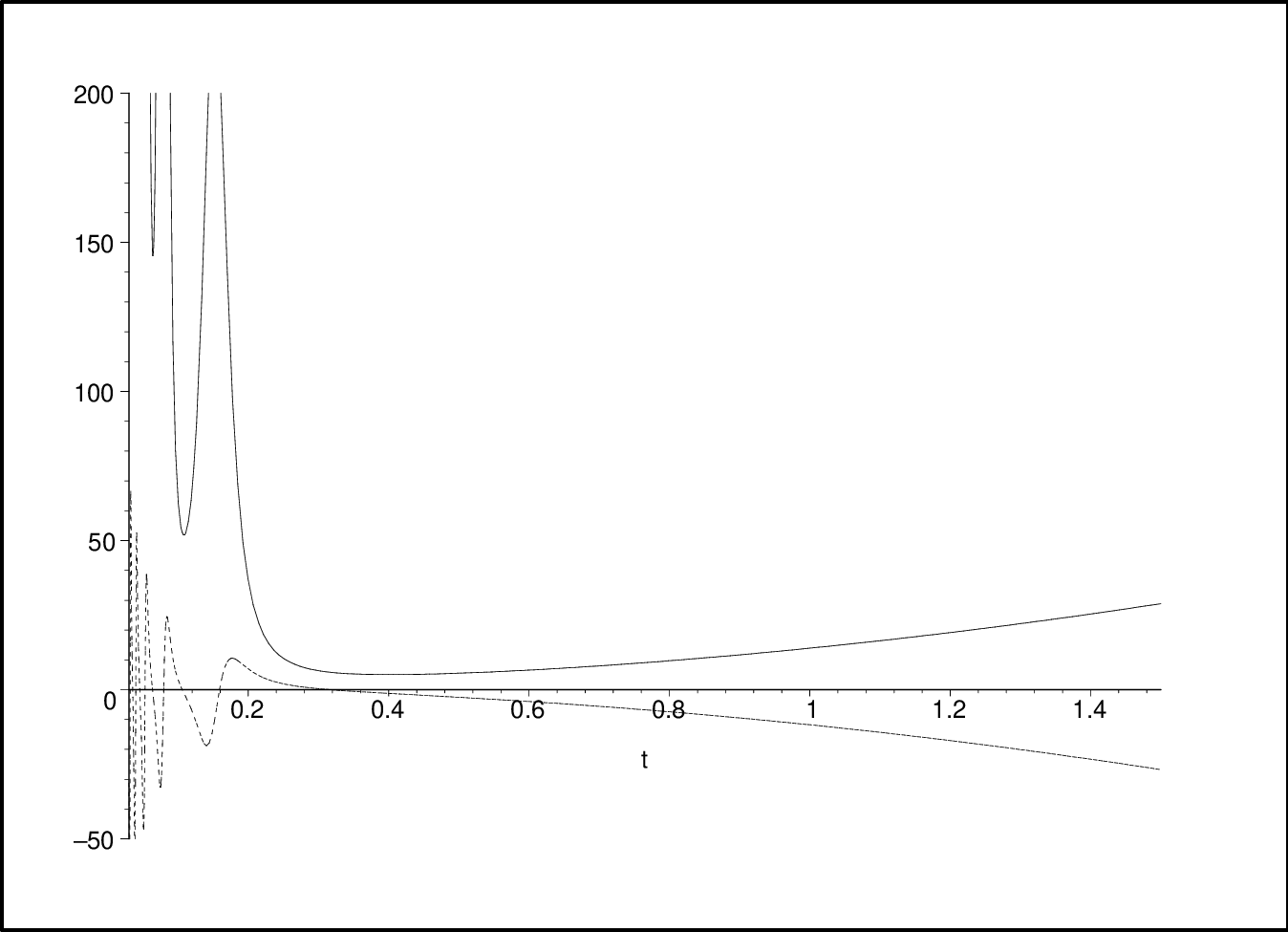}
	\caption{$\rho_6$ (solid line) and $p_6$ (dotted line) as functions of $t$.}
	\label{fig:5a}
\end{figure}
%%%%%%%%%%%%%%%%%%%%%%%%

%%%%%% Fig. 16 %%%%%%%%%
\begin{figure}[t]
	\centering
		\includegraphics[width=0.3 \textwidth, angle=270]{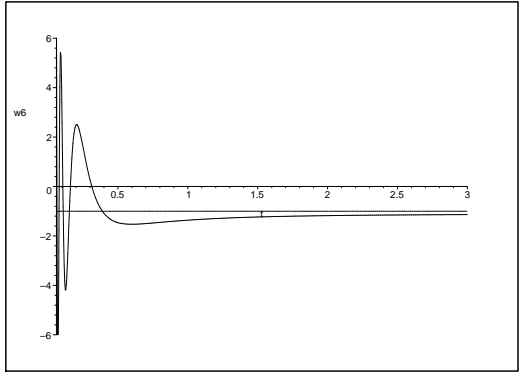}
	\caption{$w_6$ as a function of $t$.}
	\label{fig:5b}
\end{figure}
%%%%%%%%%%%%%%%%%%%%%%%%

\begin{figure}[h]
	\centering
		\includegraphics[width=0.3 \textwidth, angle=270]{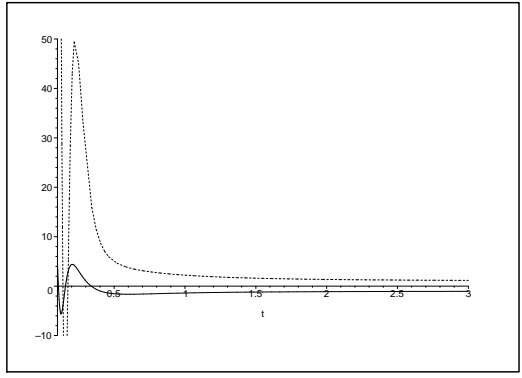}
	\caption{$q_6$ (solid line) and $j_6$ (dotted line) as functions of $t$}
	\label{fig:6j}
\end{figure}

In Fig.~\ref{fig:5a}, we display the cosmological evolution of the energy density $\rho_6$ and pressure $p_6$ as functions of $t$. 
In Fig.~\ref{fig:5b}, we show the cosmological evolution of the EoS $w_6$ as a function of $t$. 
It follows from Fig.~\ref{fig:5b} that 
in this case the crossing of the phantom divide can occur. 
Furthermore, the jerk parameter is expressed as 
\begin{align}
j_6=\frac{1}{16t^2}[18+97t^2+9t\sin[\frac{2}{t}]+18t(3t^3\sin[\frac{1}{t}]-2t^2-3)\sin[\frac{1}{t}]-\notag\\
	-9(\cos[\frac{1}{t}](7t^2+1)+4t^2+2)\cos[\frac{1}{t}]]\,.
\end{align}
In Fig.~\ref{fig:6j}, we demonstrate the cosmological evolutions of the deceleration parameter $q_6$ and jerk parameter $j_6$ as functions of $t$.

%%%%%% Fig. 17 %%%%%%%%%
%%%%%%%%%%%%%%%%%%%%%%%%

%%%%%%%%%%%%%%%%%%%%%%%%%%%
%%%  Sec. IV B 3
%%%%%%%%%%%%%%%%%%%%%%%%%%%
\subsubsection{Example 3}

We express the scale factor as 
\begin{equation}
a=a_0+b\sin [kt]+d\cos[kt]\,,
\end{equation}
where $a_0$, $b$, $d$ and $k$ are constants. {}From this description, we obtain
\begin{eqnarray}
	H&=&\frac{k(b\cos [kt] -d\sin [kt])}{a_0+b\sin [kt]+d\cos [kt]}\,,\\
	u&=&\frac{c}{(a_0+b\sin [kt]+d\cos [kt])^3}\,, \\
	\dot{\phi}&=&\frac{\phi_0}{(a_0+b\sin [kt]+d\cos [kt])^3}\,,\\
\psi_l&=&\frac{c_l}{(a_0+b\sin [kt]+d\cos[kt])^{1.5}}e^{iD}\,,\\
\psi_k&=&\frac{c_k}{(a_0+b\sin [kt]+d\cos[kt])^{1.5}}e^{-iD}\,,
\end{eqnarray}
where 
\begin{align}
D=\frac{1}{c\ \sigma}\int(2 k^2(a_0+b\sin [kt]+d\cos [kt])(a_0(b\sin[kt]+d\cos[kt])+b^2+d^2)+\notag\\+\phi^2_0 (a_0+b\sin [kt]+d\cos [kt])^{-3})dt\,.
\end{align}
The corresponding potential is given by 
\begin{equation}
V_2=-\frac{0.5\epsilon\phi_0^2}{(a_0+b\sin [kt]+d\cos [kt])^6}+3k^2\left(\frac{b\cos [kt] -d\sin [kt]}{a_0+b\sin [kt]+d\cos [kt]}\right)^2\,. 
\end{equation}
The energy density and the pressure have the forms
%\begin{equation} 
\begin{eqnarray}
\label{rho1}
\rho_7 \Eqn{=} 3k^2\left(\frac{b\cos [kt] -d\sin [kt]}{a_0+b\sin [kt]+d\cos [kt]}\right)^2\,, \\
%\end{equation}
%\begin{equation}
p_7 \Eqn{=} -k^2\frac{3(b\cos [kt]-d\sin [kt])^2-2[a_0(b\sin [kt]+d\cos [kt])+b^2+d^2]}{(a_0+b\sin [kt]+d\cos [kt])^2}\,.
\end{eqnarray}
%\end{equation}
In this example, the EoS and the deceleration parameter are described as 
\begin{equation}
	w_7=-1+\frac{2[a_0(b\sin [kt]+d\cos [kt])+b^2+d^2]}{3(b\cos [kt]-d\sin [kt])^2}
\end{equation}
and 
\begin{equation}
	q_7=-1+\frac{[a_0(b\sin [kt]+d\cos [kt])+b^2+d^2]}{(b\cos [kt]-d\sin [kt])^2}\,,
\end{equation}
respectively.

%%%%%% Fig. 18 %%%%%%%%%
\begin{figure}[t]
	\centering
		\includegraphics[width=0.3 \textwidth, angle=270]{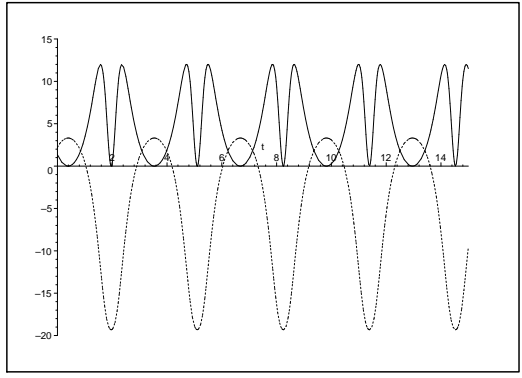}
	\caption{$\rho_7$ (solid line) and $p_7$ (dotted line) as a functions of $t$ for $a_0=2$, $b=1$, $d=1$ and $k=2$.}
	\label{fig:6a}
\end{figure}
%%%%%%%%%%%%%%%%%%%%%%%%

%%%%%% Fig. 19 %%%%%%%%%
\begin{figure}[t]
	\centering
		\includegraphics[width=0.3 \textwidth, angle=270]{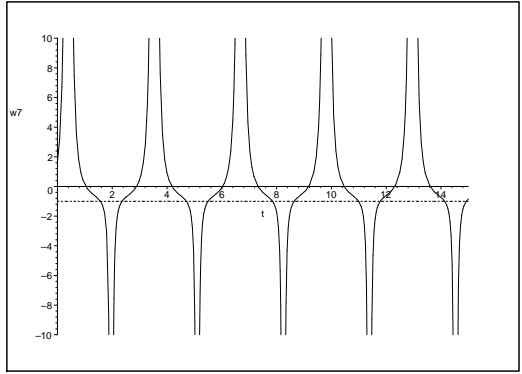}
	\caption{$w_7$ as a function of $t$ for $a_0=2$, $b=$, $d=1$ and $k=2$.}
	\label{fig:6b}
\end{figure}
%%%%%%%%%%%%%%%%%%%%%%%%

\begin{figure}[t]
	\centering
		\includegraphics[width=0.3 \textwidth, angle=270]{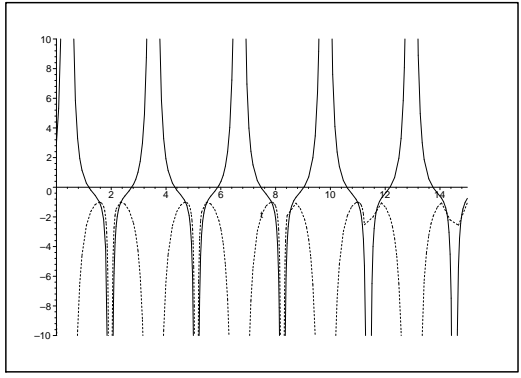}
	\caption{$q_2$ (solid line) and $j_2$ (dotted line) as functions of $t$.}
	\label{fig:7j}
\end{figure}

In Fig.~\ref{fig:6a}, we show the cosmological evolutions of the energy density $\rho_7$ and pressure $p_7$ as the functions of $t$. 
In Fig.~\ref{fig:6b}, we illustrate the cosmological evolution of the EoS $w_7$ as a function of $t$. We find from Fig.~\ref{fig:6b} that 
in this model the crossing of the phantom divide can happen. 
Furthermore, the jerk parameter is written by 
\begin{align}
	j_7=-\frac{(a_0+b\sin [kt]+d\cos [kt])^2}{(b^2-d^2)\cos^2[kt]-bd\sin[2kt]+d^2}\,.
\end{align}
In Fig.~\ref{fig:7j}, we display the cosmological evolutions of the deceleration parameter $q_7$ and jerk parameter $j_7$ as functions of $t$.

%%%%%% Fig. 20 %%%%%%%%%
%%%%%%%%%%%%%%%%%%%%%%%%

%%%%%%%%%%%%%%%%%%%%%%%%%%%
%%%  Sec. IV B 4
%%%%%%%%%%%%%%%%%%%%%%%%%%%
\subsubsection{Example 4}

We start from the following expression for the scale factor
\begin{equation}
a=\e^{\dfrac{2}{3}\int\dfrac{dt}{t+\int(l f^{-2}+m f^{-1}+n)dt}}\,,         
\end{equation}
where $f=a_0+b\cosh[kt]+d\sinh[kt]$, and $l$, $m$, $n$, $a_0$, $b$, $d$ and $k$ are constants. Then, we obtain
\begin{eqnarray}
H&=&\dfrac{2}{3[t+\int(l f^{-2}+m f^{-1}+n)dt]}\,,\\
u&=&c\ \e^{-2\int\dfrac{dt}{t+\int(l f^{-2}+m f^{-1}+n)dt}}\,, \\
\dot{\phi}&=&\phi_0 \e^{-2\int\dfrac{dt}{t+\int(l f^{-2}+m f^{-1}+n)dt}}\,,\\
\psi_l&=&c_l \e^{iD-\int\dfrac{dt}{t+\int(l f^{-2}+m f^{-1}+n)dt}}\,,\\
\psi_k&=&c_k \e^{-iD-\int\dfrac{dt}{t+\int(l f^{-2}+m f^{-1}+n)dt}}\,,
\end{eqnarray}
where 
\begin{align}
D=\frac{1}{c\ \sigma}\int(-\dfrac{4(1+l f^{-2}+m f^{-1}+n)}{3(t+\int(l f^{-2}+m f^{-1}+n)dt)^2}\ \e^{2\int\dfrac{dt}{t+\int(l f^{-2}+m f^{-1}+n)dt}}+\notag\\+\phi^2_0\int \e^{-2\int\dfrac{dt}{t+\int(l f^{-2}+m f^{-1}+n)dt}})dt\,.
\end{align}
The corresponding potential becomes
\begin{equation}
V_2=-0.5\epsilon\phi_0^2 \e^{-4\int\dfrac{dt}{t+\int(l f^{-2}+m f^{-1}+n)dt}}+\dfrac{4}{3[t+\int(l f^{-2}+m f^{-1}+n)dt]^2}.
\end{equation}
The energy density and pressure are given by 
%\begin{equation} 
\begin{eqnarray}
\rho_8=\dfrac{4}{3[t+\int(l f^{-2}+m f^{-1}+n)dt]^2}\,, 
\label{rho1} \\ 
%\end{equation}
%\begin{equation}
p_8=-\dfrac{4(l f^{-2}+m f^{-1}+n)}{3[t+\int(l f^{-2}+m f^{-1}+n)dt]^2}\,.
\end{eqnarray}
%\end{equation}
In this example, the EoS and the deceleration parameter have 
the following forms 
\begin{equation}
	w_8=l f^{-2}+m f^{-1}+n
\end{equation}
and 
\begin{equation}
	q_8=0.5+1.5(l f^{-2}+m f^{-1}+n)\,,
\end{equation}
or
\begin{equation}
	w_8=l (a_0+b\cosh[kt]+d\sinh[kt])^{-2}+m (a_0+b\cosh[kt]+d\sinh[kt])^{-1}+n
\end{equation}
and 
\begin{equation}
	q_8=0.5+1.5(l (a_0+b\cosh[kt]+d\sinh[kt])^{-2}+m (a_0+b\cosh[kt]+d\sinh[kt])^{-1}+n)\,.
\end{equation}

\begin{figure}[h]
	\centering
		\includegraphics[width=0.3 \textwidth, angle=270]{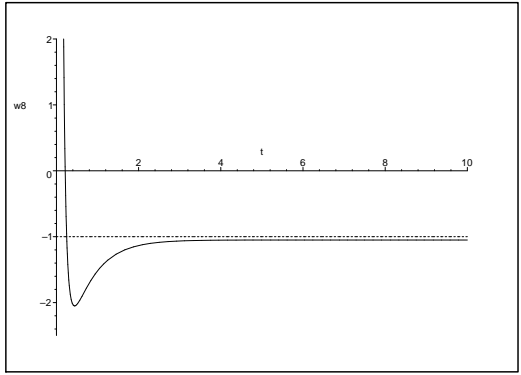}
	\caption{$w_8$ as a function of $t$ for $l=1$, $m=-2$, $n=-1$, $a_0=-1/2$, $b=1/2$, $d=1$ and $k=\sqrt{3}$.}
	\label{fig:8}
\end{figure}

In Fig.~\ref{fig:8}, we illustrate the EoS $w_8$ as a function of $t$. 
{}From Fig.~\ref{fig:8}, we understand that 
in this case the crossing of the phantom divide can be realized from 
the non-phantom (quintessence) phase (in which $w > -1$) 
to the phantom one (in which $w<-1$). 
Moreover, the jerk parameter is described by 
\begin{align}
	j_8=\frac{1}{4f^4}[18(f^2l+f^3m+l^2)+36lf(m+n)+18mf^2(m+2fn) +2f^4(2+9n^2)+\notag\\ +9f\dot{f}(2l+fm)[t+\int(l f^{-2}+m f^{-1}+n)dt]]\,.
\end{align}

%%%%%% Fig. 21 %%%%%%%%%
%%%%%%%%%%%%%%%%%%%%%%%%

As a consequence, 
from Figs.~\ref{fig:1b}, \ref{fig:3b}, \ref{fig:4b} and \ref{fig:7b} 
we understand that 
in the above four models, the universe is always in the non-phantom 
(quintessence) phase and therefore 
the crossing of the phantom divide cannot be realized. 
In particular, in the first three examples, the phantom crossing from 
the non-phantom phase to the phantom one and that from the phantom phase 
to the non-phantom one can occur. 
Moreover, in the last example, 
the crossing of the phantom divide can be realized from 
the non-phantom (quintessence) phase 
to the phantom one (in which $w<-1$). 
Such a transition behavior is consistent with the observational suggestions~\cite{Alam:2003fg, Alam:2004jy, Alam:2006kj, Nesseris:2006er, Wu:2006bb, Jassal:2006gf}. 
%%%
We also note that 
the qualitative behaviors of the numerical results shown in Figs.~\ref{fig:1a}--\ref{fig:8} do not strongly depend on the initial conditions as well as the model parameters. Presumably, from such a nature it seems that 
even if we include matter, the matter-dominated stage would be realized and 
an attractor solution could be realized. 
%%%

%%%%%%%%
We caution that the solutions reconstructed in this work might well be 
unstable. 
In addition, 
there appears divergence in the effective energy density 
in most of the examples during finite time. 
The stability of those models will be a crucial problem, which 
will spoil the models even if matter sectors are included. 
Accordingly, in order to solve this stability issue, 
we have to extend the present toy models to be more elaborate ones. 
Indeed, the resultant solutions could correspond to 
more complex models such as theories with non-linear 
kinetic terms, e.g., the ghost condensate scenario~\cite{ArkaniHamed:2003uy} and the Galileon gravity models~\cite{Nicolis:2008in, Deffayet:2009wt, Deffayet:2009mn, Deffayet:2010zh, Shirai:2012iw}. 
It would be the most important future work of the present paper 
to examine the stability of the obtained solutions 
by comparing with the above theories which consist of non-linear 
kinetic terms, so that we can find some clue for the mechanism 
to make the derived cosmological solutions viable. 
%%%%%%%%

%%%%%%%%
Consequently, the divergence or oscillating behavior 
of effective energy density will dramatically change the history of
the universe, so that the process of the nucleosynthesis will probably
be spoiled. Thus, it is crucial task for the present models 
to preserve the big bang nucleosynthesis (BBN), so that 
g-essence models can be realistic ones which are able to successfully resolve 
the mechanism of the current accelerated expansion of the universe. 
%%%%%%%%

%%%%%%%%
Furthermore, we also note that 
the reconstructed expression of the potential $V_2$ is given by a function of 
$t$ and that of not $\phi$. This is because it is difficult to obtain the 
explicit analytic form of $V_2=V_2(\bar\psi\psi)$ and we find the analytic representation of $\dot{\phi}$ and that of not $\phi$. In this respect, the reconstructed procedure performed in this work would not perfect in order to derive the form of a g-essence model as a scalar field theory and investigate the effective mass through the potential expression. 
Since we concentrate on the cosmological evolution of the EoS for 
dark energy, the reconstruction of explicit form of the potential is an additional result of this work. 
%%%%%%%%
%%%%%
%%%%%%%%
In fact, it should be noted that what 
$V_2(\bar\psi\psi)$ is given as a solution of the equations of motion 
is very peculiar. 
In this work, we consider a toy model of g-essence 
in terms of the fermion $\psi$ 
phenomenologically. However, if we take a concrete fundamental theory 
in particle physics, it is considered that 
the particle physics theory can present the form of such kind of 
effective potential for $\psi$.  
Furthermore, this situation would be similar to that 
in the quintessence models 
on the level that the potential cannot be given by particle physics 
because we do not have any particle physics theory which can 
give the very small mass of quintessence to realize 
the current cosmic acceleration, namely, 
in order for the current value of the quintessence potential to be 
equal to the very small value of the cosmological 
constant at the present universe. 
%%%%%%%%

%%%%%%%%
Finally, 
it is meaningful to emphasize the novel ingredients of this work. 
In g-essence models, there exist two dynamical components. One is the scalar 
field $\phi$. Another is the fermion condensate $u=\bar{\psi}\psi$. These two 
components play a role of dark energy. In our analysis, the dynamics of both 
$\phi$ and $u$ is included. 
As a consequence, the evolutions of cosmological quantities such as 
the deceleration parameter, the jerk parameter and the EoS in 
g-essence models become different form those in 
two scalar field models. This is one of the most important results obtained 
in this work. 
In addition, in Sec.~IV B 
we explicitly illustrate that there exist g-essence models 
in which the crossing of the phantom divide can be realized. 
In general, from the quantum field theoretical view, in a single scalar 
field theory, 
if the EoS of the scalar field is less than $-1$, i.e., the universe is in the 
phantom phase, then the null energy condition is violated, so that 
the vacuum can be unstable, and hence there will appear microphysical ghosts. 
However, in g-essence models, in the field equation level there exist 
couplings of the fermion fields to $H$, and thus the dynamics of all 
the fields in the basic system is not exactly equivalent to that in 
a single scalar field theory. Accordingly, it is not so trivial whether 
the ghosts will exist in g-essence models examined in Sec.~IV B and 
the vacuum will be unstable. 
The investigations on 
the existence of the ghosts as well as the instabilities of the vacuum 
such as Laplace instabilities 
should be the future work in our studies of g-essence models with 
realizing the crossing of the phantom divide. 
The main purpose of this paper is to illustrate that 
the crossing of the phantom divide can occur in 
the framework of g-essence models, 
although there have been proposed other simpler models 
showing the crossing of the phantom divide in the literature. 
%%%%%%%%

%%%%%%%%%%%%%%%%%%%%%%%%%%%
%%%  Sec. V
%%%%%%%%%%%%%%%%%%%%%%%%%%%
\section{Conclusions}  
  
In the present paper, we have studied the cosmological EoS for dark energy 
in g-essence models. 
In particular, we have found several g-essence models in which 
the universe is always in the non-phantom 
(quintessence) phase and thus 
the crossing of the phantom divide cannot occur. 
Furthermore, we have examined an explicit g-essence model in which 
the crossing of the phantom divide can be realized from 
the non-phantom phase to the phantom one. This transition manner is 
compatible with the analyses of the recent various cosmological observational 
data~\cite{Alam:2003fg, Alam:2004jy, Alam:2006kj, Nesseris:2006er, Wu:2006bb, Jassal:2006gf}. 

%%%%%
We remark that 
in Ref.~\cite{Komatsu:2010fb}, 
the limit on a constant EoS for dark energy has been analyzed as 
$w_{\mathrm{DE}} = -1.10 \pm 0.14 \, 
(68 \% \, \mathrm{CL})$ in a flat universe. 
Moreover, 
in case of a time-dependent EoS for dark energy 
obeying a linear form $w_{\mathrm{DE}}(a) = w_{\mathrm{DE}\,0} + 
w_{\mathrm{DE}\,a} \left( 1-a \right)$~\cite{Chevallier:2000qy, Linder:2002et} 
with $w_{\mathrm{DE}\,0}$ and $w_{\mathrm{DE}\,a}$ 
being the current value of $w_{\mathrm{DE}}$ and 
its derivative, respectively, 
the constraints have been estimated as 
$w_{\mathrm{DE}\,0} = -0.93 \pm 0.13$ and 
$w_{\mathrm{DE}\,a} = -0.41^{+0.72}_{-0.71} \, (68 \% \, \mathrm{CL})$~\cite{Komatsu:2010fb}. 
%%%%%
If the more precise future CMB experiments 
such as PLANCK~\cite{Planck-HP}
%Planck-1, Planck-2} 
support the phantom phase, 
the studies on the crossing of the phantom divide become more important 
in order to investigate the nature and property of dark energy. 
Thus, it can be considered that our results would be worthy of a clue to obtain the significant features of dark energy.

%%%%%%%%%%%%%%%%%%%%%%%%
%%%  Acknowledgments
%%%%%%%%%%%%%%%%%%%%%%%%
\section*{Acknowledgments}

%%%
K.B. would like to sincerely 
appreciate very kind and warm hospitality 
at Eurasian National University very much, where this work has been executed.

\end{document}